# Magnetically tunable dielectric materials

G. Lawes[a], T. Kimura[b], C.M. Varma[c], M.A. Subramanian[d], R.J. Cava[e], and A.P. Ramirez[f]

[a]Department of Physics and Astronomy, Wayne State University, Detroit, MI 48201
Fax: 1-313-577-3932, e-mail: glawes@wayne.edu
[b]Los Alamos National Laboratory, Los Alamos, NM 87545
Fax: 1-505-665-7652, e-mail: tkimura@lanl.gov
[c]Department of Physics, University of California Riverside, Riverside CA 92520
Fax: 1-951-827-4529, email: varma@physics.ucr.edu
[d]Dupont Central Research and Development, Experimental Station, Wilmington, DE 19880-0328
Fax:1-302-695-9799   email: mas.subramanian@usa.dupont.com
[e]Department of Chemistry, Princeton University, Princeton, NJ 08544
Fax: 1-609-258-6746, email: rcava@princeton.edu
[f]Bell Laboratories, Lucent Technologies, Murray Hill, NJ 07974
Fax: 1-908-582-4868,e-mail: apr@lucent.com

ABSTRACT: The coupling between localized spins and phonons can lead to shifts in the dielectric constant of insulating materials at magnetic ordering transitions. Studies on isostructural $SeCuO_3$ (ferromagnetic) and $TeCuO_3$ (antiferromagnetic) illustrate how the q-dependent spin-spin correlation function couples to phonon frequencies leading to a shift in the dielectric constant. A model is discussed for this spin-phonon coupling. The magnetodielectric coupling in multiferroic materials can be very large at a ferroelectric transition temperature. This coupling is investigated in the recently identified multiferroic $Ni_3V_2O_8$.

## 1. INTRODUCTION

There is considerable interest in understanding systems in which the dielectric and magnetic properties are coupled. Materials exhibiting large magnetodielectric couplings could be used in fabricating new devices, including tunable filters, magnetic sensors, and spin-charge transducers[1]. There are two general approaches to developing magnetodielectric materials. The first is to combine magnetic and dielectric materials on the mesoscale, so a magnetodielectric coupling arises from macroscopic elastic deformations[2]. The second approach is to develop single-phase materials with intrinsic magnetodielectric coupling. In the following discussion, we consider only such intrinsic magnetodielectrics.

The dielectric response of insulators is largely determined by optical phonon frequencies, as expressed by the Lyddane-Sachs-Teller relation, but these lattice distortions are normally insensitive to magnetic order. However, in certain systems, large spin-phonon coupling can produce substantial magnetization-induced shifts in the static dielectric constant. In these systems, the dielectric constant often shows a pronounced anomaly at a magnetic ordering transition. Magnetic order induced shifts in the dielectric constant have been observed in $BaMnF_4$ [3], $EuTiO_3$ [4], and $BiMnO_3$ [5] among other materials.

In this manuscript, we will address some of the questions surrounding the origin of magnetodielectric coupling in these materials. In particular we will discuss the role of ferromagnetic and antiferromagnetic correlations on the dielectric constant, and also discuss materials in which this coupling is sufficiently large to trigger a magnetically induced ferroelectric phase transition. One of the principle results of this work is that the magnetodielectric coupling depends very strongly on the symmetries of the magnetically ordered phase. In general, the dielectric constant can be more strongly coupled antiferromagnetic order having large magnetic unit cells.

## 2. EXPERIMENTAL

We investigated several different samples over the course of this investigation. The $SeCuO_3$ and $TeCuO_3$ powder samples were prepared using a solid state reaction at 700 C under 60 kbar pressure, using $SeO_2$, $TeO_2$, and CuO starting materials. The $Ni_3V_2O_8$ single crystal was grown from a $BaO$-$V_2O_5$ flux.

The magnetic properties of these materials were measured using a Quantum Design MPMS SQUID magnetometer. The magnetization was measured both as a function of magnetic field at fixed temperature, and as a function of temperature at fixed field. This allowed detailed comparisons to be made between the magnetic and dielectric properties. In order to measure the dielectric constant, gold electrodes were evaporated on opposite, parallel faces of the sample, and thin Pt wires were attached to these electrodes using conducting silver epoxy. The sample capacitance was measured using an Agilent 4284A LCR meter with the sample mounted in a Quantum Design PPMS to control temperature and magnetic field.

## 3. SPIN-PHONON COUPLING

In order to investigate the coupling between the dielectric constant and magnetic order, we studied the temperature dependence of the dielectric constant of SeCuO3 and TeCuO3. These are isostructural compounds, having a distorted perovskite lattice due to the small $Se^{4+}$ and $Te^{4+}$ ions[6]. The magnetic properties arise entirely from the $Cu^{2+}$ ions. $SeCuO_3$ is ferromagnetic, with a transition temperature of $T_c$=25 K, while the Cu-O-Cu bond angle in TeCuO$_3$ is slightly larger, leading to antiferromagnetic interactions and a transition temperature of $T_N$=9 K [6].

Both SeCuO$_3$ and TeCuO$_3$ are insulating, which allows us to measure their dielectric properties. The dielectric constants of SeCuO$_3$ and TeCuO$_3$ at fixed magnetic fields are plotted as a function of temperature in Fig. 1. It should be emphasized that because the magnitudes of the dielectric constants for SeCuO$_3$ and TeCuO$_3$ are very similar, we believe that the intrinsic dielectric properties of these materials are almost identical, and attribute any deviations to the influence of different magnetic structures in these two systems.

There are two main features in the magnetodielectric coupling illustrated in Fig. 1 which need to be explained: 1) The dielectric constant of SeCuO$_3$ drops sharply below the ferromagnetic transition temperature, and 2) The dielectric constant of TeCuO$_3$ shows a gradual increase at temperatures well above the antiferromagnetic transitions temperature, while that of SeCuO$_3$ does not.

As a first approximation, the magnetodielectric effect can be motivated as a coupling between the uniform polarization (P) and net magnetization (M) producing a term in the free energy like F=$\alpha$P$^2$M$^2$, (with $\alpha$ a constant). This is the simplest coupling between P and M that leads to a scalar free energy. This expression has been used previously to interpret the magnetodielectric coupling in EuTiO$_3$ [4], and is sufficient to understand the dielectric features observed in SeCuO$_3$ shown in Fig. 1.

At high temperatures, the net magnetization of SeCuO$_3$ is approximately zero, so there is no shift in the dielectric constant. Below $T_c$, a spontaneous magnetization develops, which couples to the electric polarization and changes the dielectric constant. As a magnetic field is applied to the sample, the decrease in dielectric constant shifts to higher temperatures and shows broadening, consistent with the behavior of the magnetization in the ferromagnetic materials. Detailed investigations show that the change in dielectric constant in SeCuO$_3$ can be fit by a term proportional to M$^2$ [7]. However, this simple coupling between uniform polarization and the net magnetization is unable to account for the observed magnetodielectric shift in TeCuO$_3$, where M would always be approximately zero.

In order to explain the magnetodielectric coupling in antiferromagnets, we have proposed a model[7] coupling the uniform polarization (P) to the q-dependent magnetic correlation function <M$_q$M$_{-q}$>, leading to a term in the free energy like:

$$F=\Sigma_q g(q)P^2 <M_q M_{-q}>. \qquad (1)$$

This extension to the previous model is sufficient to account for antiferromagnetic order; while M is zero for an antiferromagnet, <M$_q$M$_{-q}$> will develop a peak near the magnetic Bragg-vector at the zone boundary. The q-dependent coupling function g(q) can be determined by expanding the magnetic exchange integral in terms of atomic displacements. Physically, this corresponds to a coupling between the magnetic correlation function and phonon frequencies, which in turn determine the dielectric constant of the material. It can be shown that g(q) vanishes at q=0, and takes on it's maximum value on a the zone boundary, where <M$_q$M$_{-q}$> has a peak in antiferromagnets.

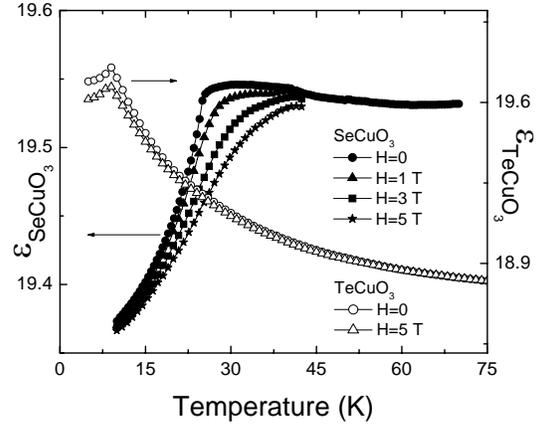

Figure 1. Temperature dependence of the dielectric constant of SeCuO$_3$ (solid symbols) and TeCuO$_3$ (open symbols) at different fixed magnetic fields.

The form of the magnetodielectric coupling in by Eq. 1 can be used to explain the temperature and magnetic field dependence of the dielectric constant of TeCuO$_3$, illustrated in Fig. 1. At high temperatures, above the antiferromagnetic transition, <M$_q$M$_{-q}$> develops a peak at a non-zero q vector, corresponding to short range antiferromagnetic correlations. Specific heat measurements suggest that TeCuO$_3$ begins to develop short-range antiferromagnetic order below approximately 60 K [6]. The coupling term g(q) will be at a maximum at this q vector, leading to an increase in the dielectric constant as <M$_q$M$_{-q}$> gets larger approaching $T_N$. In SeCuO$_3$ however, the spin-spin correlation function does not develop a peak at the maximum of g(q), so $\epsilon$(T) remains almost flat. As an aside, this model suggests that in SeCuO$_3$, as well as TeCuO$_3$, the dielectric constant in the paramagnetic phase always contains some magnetodielectric shift arising from the high temperature flat background in <M$_q$M$_{-q}$>. The "intrinsic" dielectric constant for SeCuO$_3$ is only measured when the entire spectral weight of <M$_q$M$_{-q}$> is concentrated at q=0.

Applying a large magnetic field is expected to suppress antiferromagnetic order, leading to a smaller value of <M$_q$M$_{-q}$> at the q vector for antiferromagnetic ordering. As shown in Fig. 1 the shift in dielectric constant in TeCuO$_3$ is smaller at H=5 T than what is observed for H=0 T, consistent with additional spectral weight being concentrated at the q=0 ferromagnetic spin-spin correlation.

## 4. MAGNETODIELECTRIC COUPLING AND MULTIFERROICS

The previous discussion strongly suggests that systems developing antiferromagnetic order should show larger magnetodielectric effects than simple ferromagnets. We believe that this coupling can be sufficiently strong, particularly in materials with non-collinear long wavelength magnetic structures, that it can induce ferroelectric ordering, leading to a very strong dependence of the dielectric constant on external magnetic field.

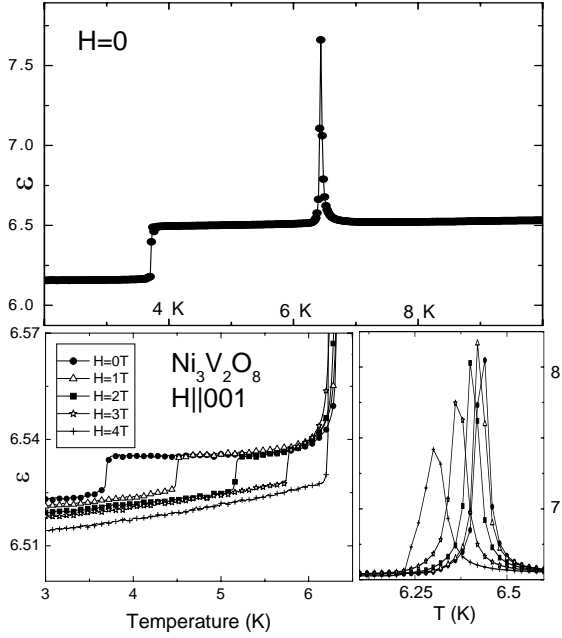

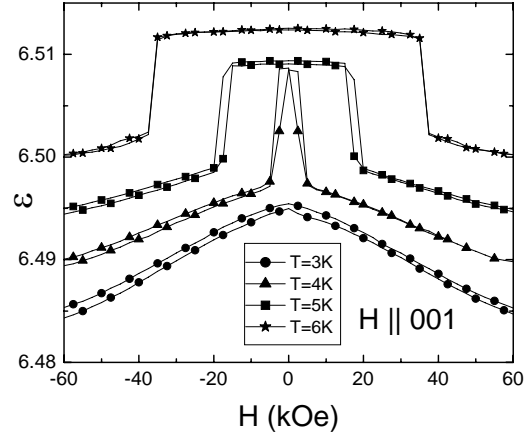

Figure 3. Dielectric constant of $Ni_3V_2O_8$ as a function of magnetic field applied along the (001) axis at different fixed temperatures.

Figure 2. Upper Panel: Dielectric constant of $Ni_3V_2O_8$ plotted versus temperature at zero magnetic field. Lower Left Panel: Temperature dependence of the dielectric constant of $Ni_3V_2O_8$ with a fixed magnetic field applied along the (001) axis. Lower Right Panel: Temperature dependence of the dielectric constant of $Ni_3V_2O_8$ at the ferroelectric phase transition. Note the change in scale.

We have investigated the magnetic and dielectric properties of $Ni_3V_2O_8$ in great detail[8-10]. $Ni_3V_2O_8$ has a layered structure, with the spin-1 $Ni^{2+}$ ions sitting at the vertices of distorted Kagome planes stacked along the (010) axis[8]. $Ni_3V_2O_8$ has a rich phase diagram: at zero magnetic field, there is a transition from the disordered phase into a high temperature incommensurate phase (HTI) at 9.3 K, followed by a transition into a low temperature incommensurate phase (LTI) at 6.3 K, and there is a canted antiferromagnetic (CAF) phase with a small ferromagnetic moment along (001) below 3.9 K. The phase diagram is sensitive to an external magnetic field; applying an external magnetic along (100) promotes the LTI phase, while a field along (001) favors the CAF phase[9].

Because $Ni_3V_2O_8$ is insulating with a small ferromagnetic moment at low temperatures, we might expect it's dielectric constant to show an anomaly at the magnetic transition temperature. This assumption is confirmed in the data presented in Fig. (2). The upper panel of Fig. (2) shows the temperature dependence of the dielectric constant at zero field, while the lower two panels show the dielectric anomalies associated with the magnetic transitions at different fields. The dielectric constant of $Ni_3V_2O_8$ shows a sharp drop coincident with the onset of canted antiferromagnetic order. This is highlighted in the lower left hand panel of Fig. (2). This can be understood using the spin-phonon model developed previously. In the CAF phase, the spin-spin correlation function develops a peak at q=0 (the contribution from the ferromagnetic canting, which is expected to be much larger than the change in $<M_q M_{-q}>$ at non-zero wavevectors), leading to a sharp drop in dielectric constant similar to that observed in $SeCuO_3$. The temperature dependence of the dielectric anomaly tracks the CAF transition as an external magnetic field is applied along the (001) axis.

In addition to the anomaly at the CAF magnetic transition, the dielectric constant of $Ni_3V_2O_8$ appears to nearly diverge at the transition between the HTI and LTI incommensurate magnetic phases, as shown the lower right hand panel of Fig. (2). This illustrates the large magnetodielectric coupling that can arise in long-wavelength non-collinear magnetic structures. Additional experiments have shown that this very large dielectric anomaly is associated with the transition into a ferroelectrically ordered state[10]. A model has been developed which explains how the complex magnetic structure of $Ni_3V_2O_8$ can couple to the uniform polarization at first order (rather than the square of the polarization arising in Eq. 1), which leads to magnetically induced ferroelectric order[10]. Similar features have been observed in other multiferroic materials, including $TbMnO_3$ [11] and $TbMn_2O_5$ [12]. The field dependence of the peak in ε(T) shown in the right panel of Fig. 2 is consistent with the shift in transition temperature from the HTI to LTI phases.

To further investigate the magnetodielectric coupling in $Ni_3V_2O_8$, we measured the dielectric constant as a function of magnetic field applied along the (001) direction at fixed temperature. These results are shown in Fig. 3. At the lowest temperature (3 K), $Ni_3V_2O_8$ remains in the canted antiferromagnetic phase over the entire range of magnetic fields. We note that the dielectric constant decreases with increasing magnetic field, consistent with observations on ferromagnetic $SeCuO_3$ discussed above. At higher temperatures, $Ni_3V_2O_8$ is in the LTI incommensurate phase at H=0, and only undergoes a transition to the CAF phase when a magnetic field is applied along the (001) direction. The sharp drop in dielectric constant at approximately H=2 kOe (T=4 K), H=17 kOe (T=5 K), and H=37 kOe (T=6 K) in Fig. 3 mark the field induced transition from the LTI phase to the CAF phase. In the LTI phase at low fields the dielectric constant is insensitive to the external magnetic field, while in the CAF phase at high fields, the dielectric constant decreases with increasing field. This is consistent with the absence of a net magnetization in the LTI phase, and the presence of a net magnetization in the CAF phase.

The sharp change in dielectric constant at this magnetic transition demonstrates the very large magnetodielectric coupling in $Ni_3V_2O_8$. Careful measurements show that at T=6 K, the dielectric constant changes by 0.12% when the applied magnetic field is changed by only 2.7% at H=18 kOe. While this is significantly smaller than the 500% magnetocapacitance observed in $DyMnO_3$ at the onset of ferroelectric order [13], this step-function decrease in dielectric constant might be more useful for developing magnetodielectric applications because of the asymmetry above and below the transition.

5. DISCUSSION

The idea that magnetodielectric effects in insulating magnets arise from spin-phonon coupling successfully explains the dielectric features observed in both ferromagnetic $SeCuO_3$ and antiferromagnetic $TeCuO_3$. In this model, it is crucial to consider the q-dependent spin-spin correlation function rather than simply the net magnetization, because otherwise the contribution from antiferromagnetic clusters will be overlooked. Experimentally, we find that the contribution from finite q spin-spin correlations can have a large effect on the dielectric properties of magnetic materials. We can heuristically motivate this by noting that at the microscopic level dielectric polarization is not a q=0 phenomena, and should be expected to couple more strongly to antiferromagnetic order.

There have been many applications proposed for ferromagnetic magnetodielectrics. Because an external field couples directly to the magnetic order parameter, it is easy to change the ferromagnetic moment and therefore produce changes in the dielectric constant. It has been suggested that because of the sensitivity of the dielectric constant to an external magnetic field these materials could be used for magnetic field sensors, dielectric actuators, and tunable filters. Antiferromagnetic magnetodielectrics on the other hand have not received as much consideration for device applications for the simple reason that it is more difficult to control the magnetic order parameter with an external field. Nevertheless, there are specific applications for which antiferromagnetic magnetodielectrics may be desirable. For devices in which the stray field arising from a ferromagnetic component would be unacceptable, antiferromagnetic magnetodielectrics could be used as non-interacting sensors, albeit possibly with a lower sensitivity.

Magnetodielectrics are expected to play an important role in the development of new technologies. While there are still numerous questions surrounding the properties of magnetodielectric materials—most notably: How can one increase the magnitude of the magnetodielectric coupling?—some of these questions are beginning to be addressed. The experimental results on $SeCuO_3$ and $TeCuO_3$ help confirm that magnetodielectric couplings can be understood in terms of spin-phonon coupling. By postulating a q-dependent term coupling the polarization to the spin-spin correlation function, we have developed a model which successfully accounted for many of the magnetodielectric features exhibited by $SeCuO_3$ and $TeCuO_3$. We also observed a large magnetodielectric coupling, predicted to be present in antiferromagnets, in the incommensurate magnetic phases of $Ni_3V_2O_8$. This system shows a large peak in dielectric constant at the transition to an incommensurate magnetic phase, and a very sharp drop in dielectric constant at the transition to a canted antiferromagnetic state.